\documentclass[prb,twocolumn,showpacs,superscriptaddress]{revtex4}
\usepackage{graphicx}
\usepackage{amssymb}
\usepackage{amsfonts}
\usepackage{amsmath}
\usepackage{lmodern}
\usepackage{mathrsfs}
\usepackage[dvipdfm,colorlinks=true, citecolor=blue, urlcolor=blue ]{hyperref}
\hypersetup{breaklinks=true}

\begin{document}

\title{Quantum phase transition in the delocalized regime of the spin-boson model}
\author{Qing-Jun Tong}
\affiliation{Center for Interdisciplinary Studies $\&$ Key Laboratory for
Magnetism and Magnetic Materials of the MoE, Lanzhou University, Lanzhou 730000, China}
\affiliation{Centre for Quantum Technologies, National University of Singapore, Singapore 117543, Singapore}
\author{Jun-Hong An}
\email{anjhong@lzu.edu.cn}
\affiliation{Center for Interdisciplinary Studies $\&$ Key Laboratory for
Magnetism and Magnetic Materials of the MoE, Lanzhou University, Lanzhou 730000, China}
\affiliation{Centre for Quantum Technologies, National University of Singapore, Singapore 117543, Singapore}
\author{Hong-Gang Luo}
\email{luohg@lzu.edu.cn}
\affiliation{Center for Interdisciplinary Studies $\&$ Key Laboratory for
Magnetism and Magnetic Materials of the MoE, Lanzhou University, Lanzhou 730000, China}
\affiliation{Beijing Computational Science Research Center, Beijing 100084, China}
\author{C. H. Oh}\email{phyohch@nus.edu.sg}
\affiliation{Centre for Quantum Technologies, National University of Singapore, Singapore 117543, Singapore}

\begin{abstract}
The existence of the delocalized-localized quantum phase transition (QPT) in the ohmic spin-boson model has been commonly recognized. While the physics in the localized regime is relatively simple, the delocalized regime shows many interesting behaviors. Here we reveal that in this regime there exists a novel QPT: namely, from a phase without a bound state to a phase with a bound state, which leads to completely different dynamical behaviors in these two phases. If the reservoir is initially in the displaced vacuum state (i.e., the coherent state), the spin dynamics exhibits lossless oscillation when the bound state exists; otherwise, the oscillation will decay completely. This result is compatible with the coherence-incoherence transition occurring in the small-tunneling limit. Our work indicates that the QPT physics in the spin-boson model needs further exploration.

\end{abstract}
\maketitle

\section{Introduction}
The spin-boson model (SBM), describing a two-level system (spin) coupled to an infinite collection of harmonic oscillators acting as the bosonic reservoir, has an extensive relevance to physical systems in condensed matter physics, quantum optics, and quantum chemistry. \cite{Weiss1999,Prior2010} An interesting aspect of the SBM is its quantum phase transition (QPT) \cite{Sachdev1999} from delocalization to localization phases with the increase of the coupling constant $\alpha$, as a consequence of the competition
between tunneling and dissipation induced by the reservoir. \cite{Leggett1987, Silbey1984, Guinea1985, Chakravarty1995, Costi1996, Stockburger1998, Bulla2003, Kopp2007, Anders2007, Zheng2007, Winter2009, Alvermann2009, Zhang2010} The QPT has been extensively studied by many different methods, including a path integral method under a non-interacting blip approximation, \cite{Leggett1987} a variational method based on unitary transformation, \cite{Silbey1984, Zheng2007} a numerical renormalization group, \cite{Bulla2003} quantum Monte Carlo simulation, \cite{Winter2009} and direct numerical diagonalization in different bases, \cite{Alvermann2009, Zhang2010} A consensus of these methods is that, for the ohmic spectrum, the delocalized-localized QPT is of Kosterlitz-Thouless type \cite{Leggett1987,Weiss1999} and occurs at $\alpha=1$ in the small-tunneling limit. Several experiments have been suggested to observe this reservoir-induced QPT; these include the use of mesoscopic metal rings, \cite{Tong2006} single-electron transistors with electromagnetic noise, \cite{Hur} and cold atoms in optical lattices. \cite{coldat}

\begin{figure}[h]
\includegraphics[width=0.9\columnwidth]{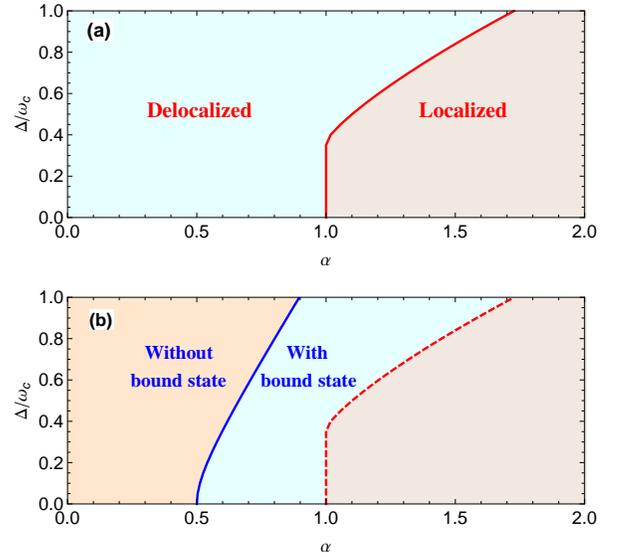}
\caption{(Color online) The phase diagram of the ohmic SBM in different parameter regimes. (a) The delocalized-localized QPT and (b) the novel QPT found in the delocalized regime, which is separated into phases with and without a bound state.} \label{QPT}
\end{figure}

While the physics in the localized regime is easy to understand, the delocalized regime exhibits many interesting phenomena. \cite{Guinea1985, Costi1996, Alvermann2009, Anders2007} For example, it was found that in the small-tunneling limit the spin dynamics in the delocalized regime shows damped coherent oscillation when $\alpha < 1/2$ and incoherent relaxation for $\alpha >1/2$, a phenomenon referred to as the coherence-incoherence transition. \cite{Leggett1987} In fact, the understanding of this dynamical transition and the physics in the regime of $1/2 < \alpha < 1$ are not clear. Here we focus on the delocalized regime and study analytically the ground-state property of the SBM by the variational method. Our starting point is to examine whether a bound state between the spin and its reservoir can form. This idea is inspired by recent experimental and theoretical developments indicating that the formation of a bound state between a two-level system and its reservoir can induce decoherence suppression of the system in a structured reservoir. \cite{Lodahl2004,Noda2007,Dreisow08,Tong2010} We find that besides the well-known delocalized-localized transition, as depicted in Fig. \ref{QPT}(a), an alternative QPT is observed in the delocalized regime, which accompanies the formation of a bound state between the spin and its reservoir, as indicated in Fig. \ref{QPT}(b). This QPT distinguishes the dynamics of the system from complete decoherence to decoherence suppression when the initial state of the reservoir is a multimode coherent state. In the small-tunneling limit this QPT happens at $\alpha = 1/2$, which is compatible with the coherence-incoherence transition point. Figure \ref{QPT} also shows the overall phase diagram for the finite tunneling amplitude. Our analytical formulation provides a clear physical picture of this QPT and a unified description of the QPT in the ohmic SBM.

Our paper is organized as follows. In Sec. \ref{model}, we introduce the SBM and its reduction by the variational approach. In Sec.
\ref{ptid}, the novel QPT occurring in the conventional delocalized-phase regime is revealed. In
Sec. \ref{dcsq} we study the dynamical consequence of this novel QPT and reveal an interesting dynamical transition at the critical point of the QPT. Finally, discussion and summary are given in Sec. \ref{das}.

\section{The SBM}\label{model}
The spin-boson model is defined as \cite{Leggett1987,Weiss1999}
\begin{equation}
H={\epsilon\over 2}\sigma_z-\frac{\Delta}{2}\sigma_{x}+\sum_{k}\omega_{k}b_{k}^{\dag}b_{k}+\sum_{k}\frac{g_{k}}{2}(b_{k}^{\dag}+b_{k})\sigma_{z},
\label{Hx}
\end{equation}
where $\epsilon$ and $\Delta$ are, respectively, the energy difference and tunneling amplitude between the two spin states, $b_{k}^{\dagger }$($b_{k}$) is the creation (annihilation) operator of the $k$th mode of the reservoir with frequency $\omega _{k}$, and $g_k$ is the coupling strength between the spin and its reservoir, which is further modeled by the spectral density
$J(\omega)=\sum_{k}g_{k}^{2}\delta(\omega-\omega_{k})=2\alpha\omega_{c}^{1-s}\omega^{s}\Theta(\omega_c-\omega)$. Here $\omega_c$ is the cutoff frequency, $\alpha$ is the dimensionless coupling constant, $\Theta(x)$ is the step function, and $s$ classifies the reservoir as subohmic if $0<s<1$, ohmic if $s=1$, and superohmic if $s>1$. \cite{Leggett1987} In the present work we focus explicitly on the ohmic spectrum, which is experimentally most relevant, and unbiased case $\epsilon=0$.

To study the ground-state property, it is convenient to make a unitary transformation $H' = UHU^\dagger$ where $U=\exp[\sum_k\lambda_k(b_k^{\dag}-b_k)\sigma_x]\exp[-\frac{i\pi}{4}\sigma_y]$ and $\lambda_k={g_k\xi_k\over 2\omega_k}$. Explicitly, $H'$ can be written as $H' = H'_{0}+H'_{1}+H'_{2}$, with
\begin{eqnarray}
&& H'_{0}=\frac{\Delta \eta}{2}\sigma _{z}+ \sum_{k}\omega_{k}b_{k}^{\dag}b_{k}-C, \nonumber \\
&&H'_{1}=\sum_{k}\nu_{k}(b_{k}^{\dag}\sigma_{-}+b_{k}\sigma_{+}),
\nonumber\\
&&H'_{2}=\frac{\Delta}{2} \sigma _{z}(\cosh \chi-\eta)-i\frac{\Delta}{2} \sigma _{y}(\sinh \chi-\eta \chi), \label{Hz}
\end{eqnarray}
where $C=\sum_{k}\frac{g_{k}^{2}\xi _{k}(2-\xi _{k})}{4\omega _{k}}$ is a constant, $\sigma_\pm$ are the raising and lowering operators in the $\sigma_z$ basis, $\chi=\sum_{k}\frac{g_{k}\xi_{k} }{\omega _{k}}(b_{k}^{\dag }-b_{k})$, $\nu_{k}=\frac{\eta\Delta g_{k}\xi_{k}}{\omega_{k}}$ acts as a renormalized coupling strength, and $\eta=\exp[-\sum_{k}\frac{g_{k}^{2}\xi_{k}^{2}}{2\omega_{k}^{2}}]$ is a renormalized factor for the tunneling. Following the variational method, \cite{Silbey1984, Zheng2007, Chin2007} the transformation parameters $\xi_{k}=\frac{\omega_{k}}{\omega_{k}+\eta\Delta}$ can be determined by minimizing the Bogoliubov-Peierls free energy. The renormalized factor $\eta$ has been used successfully to characterize the delocalized-localized QPT in the SBM. \cite{Silbey1984, Zheng2007} If the tunneling amplitude is renormalized to zero, the system is in the localized phase and the dynamics is trivial. In contrast, if the renomalized tunneling amplitude is nonzero, then the system is in the delocalized phase, which displays some interesting dynamical behaviors such as damped coherent oscillation and incoherent relaxation. \cite{Guinea1985} In Fig. \ref{QPT}(a) we reproduce the delocalized-localized transition for finite tunneling amplitude. Note that Eqs. (\ref{Hz}) take into account self-consistently the effect of counter-rotating terms in the SBM. The unitary transformation separates automatically the one-excitation and excitation-conservative transition $H'_1$ from the multiexcitation and nondiagonal transition $H'_{2}$. It has been proved that in low-temperature and weak-coupling regimes the higher-order perturbation terms $H_2'$ can be neglected. \cite{Zheng2007} Then the transformed Hamiltonian has the form $H'\approx H'_{0}+H'_{1}\equiv H_\text{eff} $.

\section{Novel QPT in the delocalized regime}\label{ptid}
It is interesting to note that $H_\text{eff}$ has the same structure as the well-known ``rotating-wave" Hamiltonian in quantum optics, \cite{Breuer2002} where the total excitation number $N=\sum_{k}b_{k}^{\dag}b_k+\sigma_+\sigma_-$ is conserved. For this kind of Hamiltonian, the Hilbert space is split into the direct sum of the subspaces with definite $N$. In this situation one can naively deem that the eigenstate $|\varphi_0\rangle=\left\vert-,\{0_k\}\right\rangle$, a tensor product of the respective ground state of the two subsystems in zero-excitation subspace with eigenvalue $E_0=-\frac12\Delta\eta$, is the ground state of the whole system. \cite{Zheng2007} Is this always true? Let us examine the eigen solution of $H_\text{eff}$. The eigenstate of $H_\text{eff}$ in single-excitation subspace can be expanded as $\left\vert\varphi_1\right\rangle=c_0\left\vert+,{0}_k\right\rangle+\sum_kc_k\left\vert-,1_k\right\rangle$, where $\left\vert1_k\right\rangle$ represents the bosonic state with only one excitation in the $k$th mode. From the eigen equation $H_\text{eff}\left\vert\varphi_1\right\rangle=E_1\left\vert\varphi_1\right\rangle$,
one can obtain a transcendental equation of the eigenvalue $E_1$,
\begin{equation}
y(E_1)\equiv\frac{\Delta\eta}{2}-\sum_k\frac{\nu_k^2}{\omega_k-(E_1+\Delta\eta/2)}=E_1,
\label{CD}
\end{equation}
If Eq. (\ref{CD}) has a real root, one can claim that the
system exhibits a bound state. From the analytic property of $y(E_1)$, one can find that Eq. (\ref{CD}) has one and only one real root in the regime $(-\infty,-{\Delta\eta\over2}]$ when $y(-{\Delta\eta\over 2})\leq-{\Delta\eta\over 2}$ is fulfilled. For the
ohmic spectrum, the condition explicitly takes the form
\begin{equation}
\alpha\geq\frac12 + \frac{\eta\Delta}{2\omega_c} \equiv\alpha_\text{c},\label{ocd}
\end{equation}
where $\alpha_\text{c}$ is the critical point of forming the bound state. It is remarkable that $E_1$ is even smaller than the eigenvalue $E_0$. Therefore, the ground-state energy is
\begin{equation} \label{grode}
E_g=\left\{ \begin{aligned}
         &E_0-C,\ \ \ \ \alpha<\alpha_\text{c} \\
         &E_1-C,\ \ \ \ \alpha>\alpha_\text{c}
                          \end{aligned} \right..
                          \end{equation}
This indicates that, with the formation of bound state, the ground state is no longer $|\varphi_0\rangle$, but $|\varphi_1\rangle$. Physically, such a sudden change of the ground-state structure manifests clearly the occurrence of QPT in the system. We can readily verify that the neglected higher-order perturbation terms $H_2'$ give no contribution in the two eigen bases, i.e. $\langle \varphi_i|H_2'|\varphi_j\rangle=0~(i,j=0,1)$, which means that the neglected term $H_2'$ has no impact on such level-crossing-caused QPT. This in turn validates our approximation. We can also verify that the eigenstates of $H_\text{eff}$ in the subspaces $N\geq2$ actually have larger eigenvalues than $E_1$. This implies that the the higher-boson states may not become the ground state.

\begin{figure}[h]
\includegraphics[width=0.95\columnwidth]{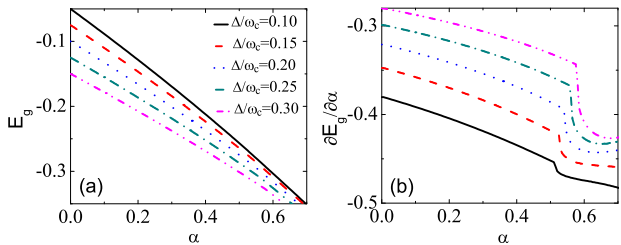}
\caption{(Color online) Ground-state energy (a) and its first derivative (b), both of which are dimensionalized to $\omega_c$, as a function of coupling constant $\alpha$. }
\label{1st}
\end{figure}
As we know, the unitary transformation does not change the eigen spectrum of the system, so the QPT occurring in $H_\text{eff}$ also means a QPT in the SBM. At zero temperature, the nonanalyticity of the ground-state energy is directly connected to the QPT. The first (or $n$th) order QPT is characterized by the discontinuity in the first (or $n$th) derivative of the ground-state energy. To further characterize this novel QPT, we plot in
Fig. \ref{1st} the ground-state energy and its first derivative as a function of $\alpha$ by numerically evaluating Eq. (\ref{grode}). We can see that, although $E_g$ is continuous, its first derivative suffers a sudden drop at the critical point $\alpha_\text{c}$. This implies that the QPT is first-order, which can also be verified by the fact that the orthogonality of the ground states at two sides of $\alpha_\text{c}$ will cause the ground-state fidelity to suddenly drop to zero. \cite{Zhang2010} We also plot in Fig. \ref{QPT}(b) the overall phase diagram of the SBM in the finite tunneling amplitude. With the increase of the bare tunneling amplitude $\Delta$, the critical point $\alpha_\text{c}$ increases. In the small-$\Delta$ limit, a lower bound $\alpha_\text{c}=\frac{1}{2}$ exists. The novel QPT occurring in the conventional delocalized phase also falls in the valid regime of our approximation, i.e. the weak-coupling regime. This also validates our approximation.

\section{Dynamical consequence}\label{dcsq}
What is the dynamical consequence of this novel QPT? Some works have shown that the formation of a bound state in a dissipative qubit system under a rotating-wave approximation is accompanied by decoherence suppression of the qubit. \cite{John1990, Tong2010} This is understandable based on the fact that the bound state, as a stationary state of the whole system, has a vanishing decay rate and the coherence contained in it would be preserved during the time evolution. In a similar manner we also expect that the dynamics in our SBM here is qualitatively changed with the occurrence of the novel QPT. In the phase without a bound state, the dynamics of the system exhibits complete decoherence, while in the phase with a bound state it exhibits decoherence suppression.

\begin{figure}[h]
\centering
\includegraphics[width=0.9\columnwidth]{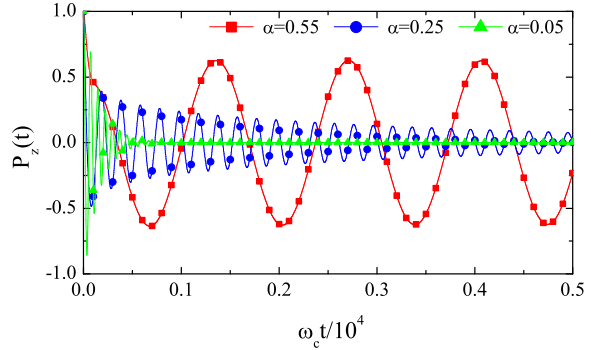}
\caption{(Color online) Time evolution of $P_z(t)$ under different coupling constants $\alpha$. The other parameter used here is $\Delta/\omega_{c}=0.1$. The cases of $\alpha=0.05$ and $0.25$ are in the phase without a bound state, and the case of $\alpha=0.55$ is in the phase with a bound state.} \label{pz}
\end{figure}
To confirm our above expectation, we study the nonequilibrium dynamics of the system, which is usually characterized by the quantity $P_z(t)=\langle \Psi(t)|\sigma_z|\Psi(t)\rangle$, with $|\Psi(t)\rangle$ being the time-dependent state of the whole system. \cite{Leggett1987,Weiss1999} We consider the initial state to be
\begin{equation}\label{int}
   |\Psi(0)\rangle=|+\rangle\otimes\prod_ke^{-\lambda_k(b_k^\dag-b_k)}|0_k\rangle,
\end{equation}
which is a product state of the excited state in the $\sigma_z$ basis of the spin and a multimode coherent state of the reservoir. It is noted that such an initial state is different from the one considered in Refs. \onlinecite{Guinea1985,Costi1996,Alvermann2009,Anders2007}. It is straightforward to verify that under the unitary transformation $U$, $P_z(t)=\langle\Psi'(t)|\sigma_x|\Psi'(t)\rangle $, where the evolution of $|\Psi'(t)\rangle\equiv U|\Psi(t)\rangle $ is governed by $H_\text{eff}$ under the initial condition $|\Psi'(0)\rangle=|+_x,\{0_k\}\rangle$ with $|+_x\rangle={|+\rangle+|-\rangle\over \sqrt{2}}$ being the eigenstate of $\sigma_x$. The evolution of the reduced density matrix $\rho'=\text{Tr}_R(|\Psi'\rangle\langle\Psi'|)$ of the spin is governed exactly by the following master equation: \cite{Breuer2002}
\begin{eqnarray}
\dot{\rho}'(t) &=&-i\frac{\Omega (t)}{2}[\sigma _{+}\sigma_-,\rho' (t)]+%
\frac{\gamma (t)}{2}[2\sigma _{-}\rho' (t)\sigma _{+}  \notag \\
&&-\sigma _{+}\sigma _{-}\rho' (t)-\rho' (t)\sigma _{+}\sigma _{-}],\label{mast}
\end{eqnarray}%
where $\Omega (t)=-2\text{Im}[\frac{\dot{c}(t)}{c(t)}]+\Delta\eta$ and $\gamma (t)=-2\text{Re}[\frac{\dot{c}(t)}{c(t)}]$. Physically, $\Omega (t)$ plays the role of time-dependent shifted frequency and $\gamma (t)$ that of time-dependent decay rate. The time-dependent factor $c(t)$ satisfies
\begin{equation}
\dot{c}(t)+i\frac{\Delta\eta}{2}c(t)+\int_0^{t}c(\tau)f(t-\tau)d\tau=0,~c(0)=1,
\end{equation}
in which the kernel function $f(x)=\int_0^{\infty}J'(\omega)e^{-i\omega x}d\omega$ connects to the renormalized spectral density
$J'(\omega)=\sum_{k}\nu^2_{k}\delta(\omega-\omega_k)$. By solving Eq. (\ref{mast}) numerically, we can get $P_z(t)=\text{Tr}_S[\rho'(t)\sigma_x]$, as shown unambiguously in Fig. \ref{pz}. We can see that $P_z(t)$ decays to zero asymptotically in an oscillatory way when the bound state is absent ($\alpha=0.25$ or $0.35$), while it shows lossless oscillation ($\alpha=0.55$) once the bound state is formed. This indicates clearly that the decoherence really can be suppressed in the phase with the bound state. This dynamical behavior in turn confirms the existence of this bound-state-induced QPT in the SBM. The sudden transition from complete decoherence to decoherence suppression is a result of QPT occurring in the model itself. It is the abrupt change of the ground-state structure near the critical point of the QPT that induces the qualitative difference in the dynamics of the SBM.

On the other hand, a widely studied case is the nonequilibrium dynamics of the spin system when the reservoir is initially in a vacuum state. \cite{Guinea1985,Costi1996,Alvermann2009,Anders2007} It was shown that the dynamics changes from damped coherent oscillation to incoherent relaxation, i.e. the so-called coherence-incoherence transition, at the point $\alpha=\frac12$ in the small-$\Delta$ limit. \cite{Leggett1987} As a dynamical behavior, this is of course not a generic phenomenon for any initial reservoir state. Therefore, the coherence-incoherence transition itself does not suffice to deduce the occurrence of the intrinsic QPT of the model. Then what is the physical reason for this dynamical transition? We notice that the transition point matches perfectly well with the critical point $\alpha_\text{c}$ in the QPT in the small-$\Delta$ limit. Thus, it is reasonable to conjecture that the coherence-incoherence transition is also a dynamical consequence of this novel QPT at the critical point $\alpha_\text{c}$.

The different dynamical consequences are actually due to the different reservoir properties. To the initial vacuum state of the reservoir, the QPT induces the coherence-incoherence dynamical transition, while to the multimode coherent state, it induces the dynamical transition from oscillatory damping to lossless oscillation. The distinctive character of the coherent state Eq. (\ref{int}) resides in that it is a vacuum state for the reservoir part in the $H_\text{eff}$ representation so that only the two lower eigen subspaces with $N=0$ and $1$ of the whole system are involved in the dynamics. This means that only one of an arbitrary infinite number of modes of the reservoir with one boson affects the interaction with the spin. As a result the reservoir shows more coherent interplay with the spin and efficiently prolongs the coherence time of the spin, especially when the bound state is formed. On the other hand, if the reservoir is initially in the vacuum state, i.e., $|\Psi(0)\rangle=|+,\{0_k\}\rangle$, the dynamical behavior of lossless oscillation can no longer appear. This is because, in this case, the reservoir in $H_\text{eff}$ representation corresponds to $\prod_k|\lambda_k\rangle$ with the total boson number being \[n=\sum_k\left\vert\lambda_k\right\vert^2=\int_0^\infty\frac{J(\omega)}{4(\omega+\Delta\eta)^2}d\omega,\] which implies that a process of a large number of bosons is involved in the dynamics. Such a large number of bosons distributed among the infinite reservoir modes inevitably induces a completely out-of-phase interaction with the spin. Consequently, the lossless oscillation cannot occur for the initial vacuum state even in the phase with the bound state. In the phase without the bound state, $\lambda_k$ and $n$ are not so large because of the small $\alpha$ in this phase. Then our results coincide qualitatively with the decaying oscillation in Ref. \onlinecite{Alvermann2009}.

\section{discussion and summary}\label{das}
In conclusion, using the variational method, we have found a novel QPT in the delocalized regime of the SBM. This QPT separates the delocalized regime into phases with and without a bound state. In these two phases, the decoherence dynamics of the spin shows qualitative differences. In particular, if the initial reservoir is in the multimode coherent state, the spin dynamics in the phase with a bound state exhibits a lossless oscillation. In contrast, in the phase without a bound state, the oscillation will decay completely to zero. Our purely analytical treatment provides a unified microscopic description of the QPT and renders a helpful understanding of the rich physics in the SBM. Further extension of the present work to other spectra is straightforward. As a final remark, the dynamics discussed in the present work only involves the zero-temperature case. It is also interesting to discuss the dynamical consequence caused by this QPT when the reservoir is initially in thermal equilibrium. \cite{Sassetti1990,Lesage1996,Egger1997} However, in this case it is very difficult to obtain the exact dynamics because the initial state involves the eigen bases of all the subspaces with definite excitation number from zero to infinity. Therefore, some approximations are necessary. This problem deserves further exploration in the future.

\section*{Acknowledgement}
This work is supported by the Fundamental Research Funds for the Central Universities, Gansu Provincial NSF of China under Grant No. 0803RJZA095, by the NSF of China under Grant Nos. 11175072 and 11174115, and by the CQT WBS Grant No. R-710-000-008-271.

\end{document}